\documentclass[12pt]{article}
\usepackage{graphicx}
\usepackage{amssymb}
\usepackage{epstopdf}
\usepackage{setspace}
\usepackage{amsmath}
\usepackage{caption}
\usepackage{gensymb}
\usepackage{physics}
\usepackage{subfig}
\usepackage{titling}
\usepackage{soul,color}
\usepackage{authblk}
\title{Tensor Network Models of \\ Unitary Black Hole Evaporation}
\author[1]{Samuel Leutheusser\thanks{sam.leutheusser@alumni.ubc.ca}}
\author[1]{Mark Van Raamsdonk\thanks{mav@phas.ubc.ca}}
\affil[1]{Department of Physics and Astronomy, University of British Columbia\\
             6224 Agricultural Road, Vancouver, British Columbia, Canada, V6T 1Z1}

\newcommand{\be}{\begin{equation}}
\newcommand{\ee}{\end{equation}}
\newcommand{\bea}{\begin{eqnarray}}
\newcommand{\eea}{\end{eqnarray}}
\newcommand{\beas}{\begin{eqnarray*}}
\newcommand{\eeas}{\end{eqnarray*}}
\newcommand{\ba}{\begin{array}}
\newcommand{\ea}{\end{array}}
\def\identity{{\rlap{1} \hskip 1.6pt \hbox{1}}}

\begin{document}

\begin{titlingpage}
\maketitle
\date{\today}

\begin{abstract}
We introduce a general class of toy models to study the quantum information-theoretic properties of black hole radiation. The models are governed by a set of isometries that specify how microstates of the black hole at a given energy evolve to entangled states of a tensor product black-hole/radiation Hilbert space. The final state of the black hole radiation is conveniently summarized by a tensor network built from these isometries. We introduce a set of quantities generalizing the Renyi entropies that provide a complete set of bipartite/multipartite entanglement measures, and give a general formula for the average of these over initial black hole states in terms of the isometries defining the model. For models where the dimension of the final tensor product radiation Hilbert space is the same as that of the space of initial black hole microstates, the entanglement structure is universal, independent of the choice of isometries. In the more general case, we find that models which best capture the ``information-free'' property of black hole horizons are those whose isometries are tensors corresponding to states of tripartite systems with maximally mixed subsystems.

\end{abstract}
\end{titlingpage}

\section{Introduction}

While semiclassical calculations \cite{creation} suggest the breakdown of unitarity in the black hole evaporation process \cite{breakdown}, it is widely believed that in a complete quantum theory of gravity, information about the initial state of the black hole can, in principle, be recovered from the final state of the Hawking radiation. For example, in the AdS/CFT correspondence, the formation and evaporation of a small black hole in Anti-de-Sitter space should be described precisely by unitary evolution of a state in the dual conformal field theory. Assuming this unitarity, it is interesting to understand more precisely the quantum-information theoretic structure of the outgoing radiation, for example to understand how deviations from thermality allow the encoding of the black hole initial state.\footnote{For a recent review on the black hole information paradox, see  \cite{Harlow}.}

In the past, various authors (see e.g. \cite{Gidd2013, MathurPed, Gidd1992, Gidd2012, GiddShi, MathurInfall, Avery, Page2016, Rozali, Bradler, Richter}) have studied toy models of black hole evaporation, in which a quantum system with Hilbert space ${\cal H}^{BH} \otimes {\cal H}^{rad}$ evolves unitarily, with information from the black hole subsystem ${\cal H}^{BH}$ being transferred somehow to the radiation subsystem ${\cal H}^{rad}$. In these models, various conditions are imposed on the dynamics in order to incorporate physical features expected of black hole evaporation. In this note, we introduce and study a very broad class of models of this type. Our goal is to make only a few plausible assumptions and provide some general tools for investigating the quantum information-theoretic structure of the resulting black hole radiation.

We impose only a few basic constraints on our models:
\begin{itemize}
\item
We require that the model is information-preserving, that is, the map between the initial black hole state and the state of the radiation subsystem after evaporation is an isometry (i.e. a unitary map to a subspace of the radiation system).
\item
In physical black hole systems, the radiation quanta have typical energies of order $k_B T$, where $T = dE/dS$ is the black hole temperature. This implies that microcanonical entropy changes by a fixed amount per quantum emitted throughout the evolution. We model this by taking the evolution to be a series of discrete steps (each corresponding to the release of one thermal particle), with the logarithm of the number of available states in the black hole subspace decreasing by the same amount at each step.
\item
We require that the emitted radiation does not influence the future evolution of the black hole. We impose this by requiring that each bit of radiation lives in a separate tensor product factor of the radiation Hilbert space, and that at each step, the current state of the black hole determines the state of the black hole plus only one of these radiation Hilbert space factors.
\end{itemize}

\begin{figure}
\centering
\includegraphics[width= 0.8 \textwidth]{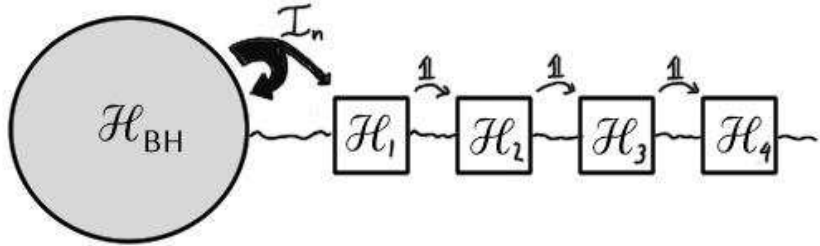}
\caption{Model for black hole evaporation. At each step, the state of the black hole is mapped via an isometry to a state of the combined system of the black hole plus the first radiation subsystem, while the state of the $n$th radiation subsystem is transferred to the state of the $(n+1)$st radiation subsystem.}
\label{BHevap}
\end{figure}

\begin{figure}
\centering
\includegraphics[width= 0.5 \textwidth]{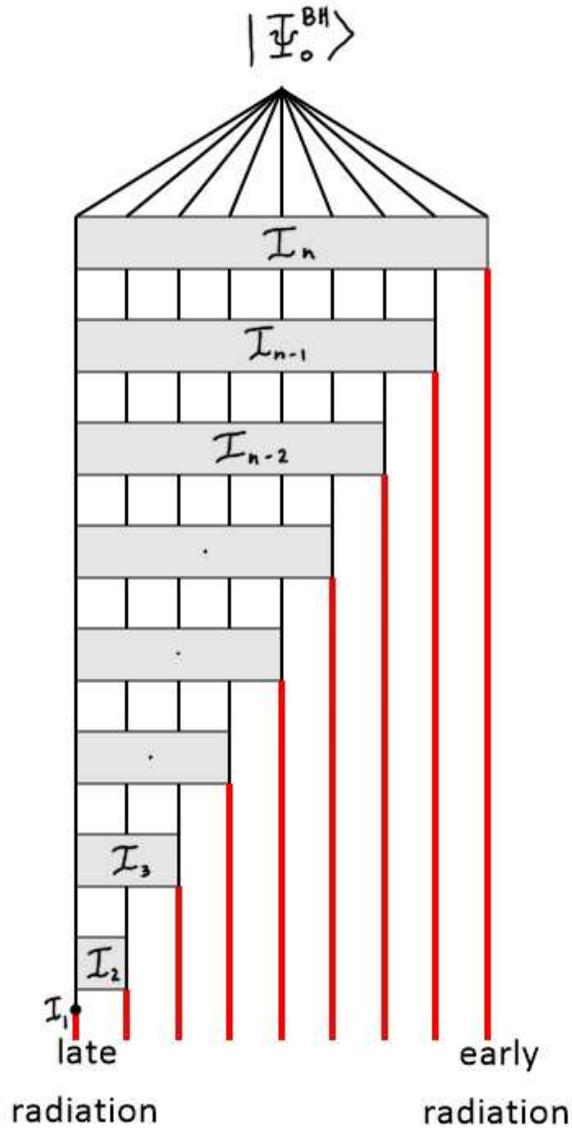}
\caption{Tensor network for final state of black hole radiation. Input state at the top is the black hole initial state, an arbitrary state of a dimension $d^n$ Hilbert space, represented as the tensor product of $n$ Hilbert spaces of dimension $d$ (thinner black lines). Each rectangular box represents an isometry. Thicker red lines correspond to $D$-dimensional tensor product factors of the radiation subsystem each containing one quantum of radiation. Maximally efficient models have $D=d$ and each isometry is a unitary.}
\label{fig_TN}
\end{figure}

In section 2 below, we describe a very general setup incorporating these constraints. We point out that the final state of the radiation subsystem containing the information about the initial black hole state is naturally described by a tensor network, as depicted in figure \ref{fig_TN}. Here, black hole microstates from an ensemble with entropy $n \log d$ are encoded in a subsystem of the radiation Hilbert space given by a tensor product of $n$ copies of a dimension $D$ Hilbert space. The information-preserving property requires that $D \ge d$, and we refer to the special case $D=d$ as `maximally efficient', since here the radiation subspace in which the black hole initial state is encoded has the same dimension as the space of black hole initial states.

In section 3, we describe various measures that can be used to understand the entanglement structure and other information-theoretic properties of the final state of the radiation. First, we can calculate entanglement entropy and R\'enyi entropies for various radiation subsystems (describing subsets of the emitted quanta) in order to characterize the basis-independent information included in the density matrix for that subsystem. We also describe a generalization of the R\'enyi entropies that captures additional information contained in the full multipartite density matrix that is independent of the choice of basis for each of the subsystems. For an $n$-part system, these  measures are calculated starting with $k$ copies of the full density matrix $\rho^{b_1 b_2 \cdots b_n}_{a_1 a_2 \cdots a_n}$, and $n$ elements $(\sigma_1, \dots , \sigma_n)$ of the permutation group $\mathfrak{S}_k$. The quantity
\be
S_k(\rho ; \sigma_1, \dots , \sigma_n)
\ee
is then defined by contracting up the $k$ pairs of indices for the $j$th subsystem according to the permutation $\sigma_j$, as explained in more detail in section 3. We argue that these quantities give a complete set of invariants under independent changes of basis for the subsystems and are general measures of multipartite entanglement in the system.

In section 4, we carry out a calculation of these entanglement invariants in a general model when we average over all the possible initial states of the black hole. We find that for a given model, there is a single density matrix  $\rho_{\cal I}$  for the tensor product radiation Hilbert space, constructed from the isometries defining the model, such that all of the averaged entanglement invariants for the model may be expressed as linear combinations of the entanglement invariants for this single ``master'' state.

In section 5, we evaluate these measures of entanglement more explicitly in the special case of the maximally efficient models. In this class of models, we show that the entanglement structure in the black hole radiation is universal: when averaged over initial states, the results for all the entanglement measures described in section 3 do not depend on the specific isometries ${\cal I}_n$ defining the model, and are simply the results for a Haar-random state of the multipart radiation system. For such systems, the entanglement entropy of a subsystem is given by the well-known result of Page \cite{Page}, and the subsystem R\'enyi entropies have been calculated previously in \cite{Malacarne, Zyc2}. We introduce new diagrammatic calculational techniques that relate these quantities and the more general quantities of section 3 to certain generating functions that appear in the theory of the symmetric group. These techniques provide an alternative derivation of the previous results, a new simpler formula for the R\'enyi-entropies, and expressions for the more general multipartite entanglement measures discussed in section 3.

In section 6, we return to the general case, and use the results of section 4 to investigate which choices of isometries give models that best capture the ``information free'' property of black hole horizons. Specifically, we determine constraints on the isometries that arise from demanding that in each radiation step, the output state in the dimension $D$ radiation subsystem is as close as possible to the maximally mixed state of that subsystem. We find that a sufficient condition is that the tensors ${\cal I}^n_{m i}$ defining the model correspond to states
\be
{\cal I}^n_{m i} |n \rangle \otimes |m \rangle \otimes |i \rangle
\ee
for which each subsystem is maximally mixed.

We end in section 7 with a discussion. Finally, some combinatoric details for the derivations of section 5 are presented in an appendix. This paper is based on the UBC undergraduate thesis \cite{SLthesis}.

\subsubsection*{Relation to earlier work}

While many of the basic features of our models have appeared before in the literature, we believe that our setup is more general than the existing models, and that the connection to tensor networks, the investigation of the broader class of entanglement measures discussed in section 3, and the calculational techniques for entanglement measures in Haar-random states are novel, revealing an interesting connection to enumerative combinatorics of the symmetric group.

\section{Tensor network models of black hole evaporation}

In this section, we introduce a general class of quantum mechanical models in which the Hilbert space takes the form ${\cal H}^{BH} \otimes {\cal H}^{rad}$, and the evolution proceeds through a series of discrete steps (corresponding to the release of radiation quanta) from an initial black hole state $|\Psi^{BH} \rangle \otimes |0^{rad} \rangle$ to a final radiation state of the form $|0^{BH} \rangle \otimes |\Psi^{rad} \rangle$.

The black hole subsystem will have states at various energies with some density of states characterized by $\rho(E) \sim exp(S_{BH}(E))$ where $S_{BH}(E)$ describes the relation between entropy and energy for the type of black hole being considered. During the evaporation process, the energy in the black hole subsystem decreases with each radiation quantum emitted. Since the energy of these quanta is of order $T$ (setting $k_B = 1$), where $T = dE/dS$ is the black hole temperature, the change in black hole entropy with each quantum emitted is of order one throughout the evaporation process:
\be
\delta S_{BH} = - {dS_{BH} \over dE} \; \delta E \sim - {1 \over T} \;  T =  {\cal O}(1) \; .
\ee
The logarithm (in some base $d$) of the number of available states therefore decreases by one at each step. To model this, we represent the black hole subsystem as a direct sum
\be
{\cal H}^{BH} = \oplus_{N=0} {\cal H}^{BH}_N
\ee
where ${\cal H}^{BH}_N$ has dimension $d^N$, and assume that a black hole state in the subsystem ${\cal H}^{BH}_N$ evolves after the emission of one quantum to a state in the lower-energy subsystem ${\cal H}^{BH}_{N-1}$.\footnote{This represents a simplification, since in a physical black hole system, there will be a spectrum of possible energies for the emitted quanta.} This state will generally be mixed, since the resulting black hole state will now be entangled with the radiation subsystem.

We will assume that once a quantum is emitted into the radiation subsystem, its subsequent dynamics is trivial and does not influence the later evaporation of the black hole. To model this, we take the radiation subsystem to be a tensor product of equivalent subsystems,
\be
{\cal H}^{rad} = {\cal H}^{rad}_1 \otimes {\cal H}^{rad}_2 \otimes \cdots
\ee
each with some dimension $D \ge d$.  A black hole state in the subspace ${\cal H}^{BH}_n$ is taken to evolve via an isometry
\be
\label{defisom}
{\cal I}_n : {\cal H}^{BH}_n \to {\cal H}^{BH}_{n-1} \otimes {\cal H}^{rad}_1
\ee
which represents the emission of a single quantum into the first radiation subsystem. The full state evolves by combining this with a map that simply transfers the state of radiation subsystem $n$ to the subsystem $n+1$. Thus, acting on an arbitrary basis element for the full system, the evolution for a single time step is
\be
|m \rangle_n \otimes |i_1 \rangle \otimes | i_2 \rangle \otimes \cdots \to \left({\cal I}_n\right)_m^{m',i} |m' \rangle_{n-1} \otimes |i \rangle \otimes | i_1 \rangle \otimes | i_2 \rangle \otimes \cdots
\ee
The isometry condition ${\cal I}_n^\dagger {\cal I}_n = \identity$ (with no summation over $n$) guarantees that the full map is information-preserving.

The model is summarized in figure \ref{BHevap}. We can think of each ${\cal H}^{rad}_n$ as the Hilbert space for a memory register in a quantum computer into which the information about each successive quantum of radiation is unitarily transferred. To make a precise connection with black hole evaporation in a complete theory of quantum gravity, we can imagine that the initial black hole is described by a high-energy state in some holographic CFT, which is coupled to some auxiliary systems as in figure \ref{BHevap}. The coupling is such that in the gravity picture, each quantum of black hole radiation is absorbed at the boundary of AdS and its quantum information is stored in the first auxiliary system, while the previously stored states are faithfully transferred along the chain.

Starting with an arbitrary state $|m \rangle_n$ in the subspace ${\cal H}^{BH}_n$ and evolving this for $n$ time steps corresponds to the complete evaporation of the black hole, which can be described as an isometry
\be
\label{genmap}
{\cal H}^{BH}_n \to {\cal H}^{rad}_1 \otimes  \cdots \otimes {\cal H}^{rad}_n
\ee
Noting that the space ${\cal H}^{BH}_n$ of dimension $d^n$ is isomorphic to the tensor product of $n$ copies of a dimension $d$ Hilbert space, we can represent the entire evolution by the tensor network shown in figure \ref{fig_TN}.

The model we have defined is information-preserving in the sense that each black hole state maps via (\ref{genmap}) onto a unique state of the final radiation system. As a special case, we can make the  choice $D=d$, which corresponds to the minimum dimension for which such an isometry is possible. In other words, information about the black hole state is encoded with maximal efficiency in the radiation system. In this case, each isometry in (\ref{defisom}) and the overall map (\ref{genmap}) are unitary transformations, since they are isometries between Hilbert spaces of the same dimension.\footnote{While the more general case with $D>d$ does not correspond to a unitary map, the important point is that they are still information-preserving. These models can still arise as the effective description of the black hole evaporation process in a unitary model. To illustrate this in a simpler example, consider a general closed quantum system (e.g. a Hydrogen atom) with Hamiltonian $H_0$ and some discrete spectrum of energy eigenstates, and couple this weakly to an open quantum system (e.g. a free electromagnetic field) which can be described by a basis of outgoing scattering states. In this case, evolution of the full system is unitary, and an initial state for which the closed system is in a general state with the open system unexcited will evolve into some final scattering state in which the closed system has relaxed to its ground state. Distinct initial states must evolve into distinct final states, but there is no requirement that the dimension of the closed system must match the dimension of the open system.}

So far, the models we are considering are quite general, and could be used to represent more general ``quantum encoder'' systems in which the information about some quantum state is stored in a quantum memory described by a tensor product Hilbert space. In section 6, we will discuss additional conditions on the isometries designed to incorporate some of the expected features specific to black hole systems. In particular, we will try to incorporate the ``information-free'' property of black hole horizons by requiring that the outgoing quantum in given radiation step contains as little information about the black hole state as possible.

\section{Measures of entanglement structure for multipartite systems}

We will be interested in the entanglement structure of the final state, which in our model lives in a tensor product Hilbert space with $n$ parts, if the initial black hole state is in ${\cal H}^{BH}_n$.
For any subsystem $A$ (a collection of $k$ of these parts) we can quantify the entanglement with the rest of the system via the entanglement entropy $S^A = - \tr( \rho_A \log \rho_A)$. More detailed information is contained in the spectrum of $\rho_A$, or equivalently in the R\'enyi entropies
\be
\label{renyi}
S^A_n \equiv \log(\tr((\rho_A)^n))/(1-n)
\ee
These quantities represent the complete basis-independent information in the reduced density matrix for the subsystem $A$.

Starting from the full density matrix
\be
\rho^{b_1 b_2 \cdots b_n}_{a_1 a_2 \cdots a_n} \; ,
\ee
there are additional quantities that we can define, generalizing the R\'enyi entropies, which are also invariant under an independent change of basis for each of the subsystems.

\begin{figure}
\centering
\includegraphics[width= 0.75 \textwidth]{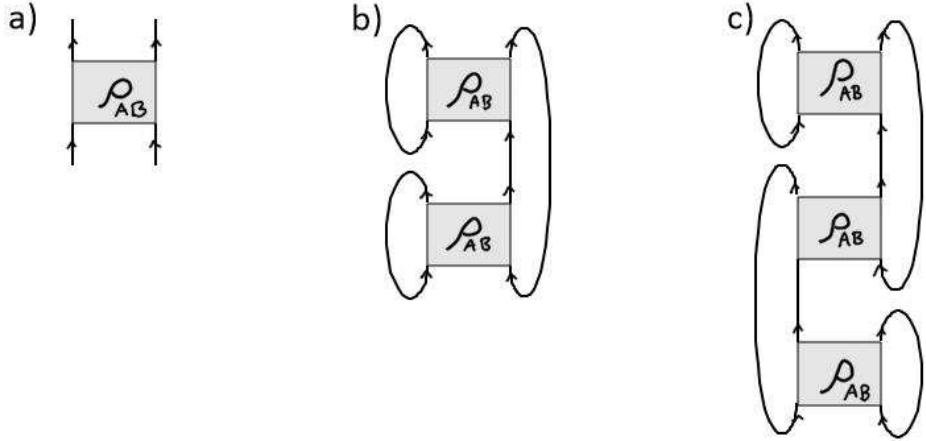}
\caption{Constructing invariants from a multipartite density matrix $\rho$: a) Diagrammatic representation of density matrix $\rho_{AB}$ for a system with two parts $A$ and $B$. Ingoing/outgoing lines represent indices transforming in the fundamental/antifundamental representation under the unitary operation corresponding to the change of basis for a subsystem. b) Diagrammatic representation of $\tr(\rho_B^2)$. c) Diagrammatic representation of $S_3(\rho_{AB} ;(1)(23),(12)(3))$}
\label{rho123}
\end{figure}

For the present discussion, the subsystems need not be identical. A change of basis in the $m$th system corresponds to a unitary transformation
\be
\rho^{a_1 a_2 \cdots a_n}_{b_1 b_2 \cdots b_n} \to U^{a_m}_{c_m} \rho^{a_1 \cdots c_m \cdots a_n}_{b_1 \cdots d_m \cdots b_n} (U^\dagger)^{d_m}_{b_m}\; .
\ee
Quantities that are invariant under all such transformations acting independently on each subsystem correspond to quantities built out of $k$ copies of the density matrix, with the $k$ lower indices corresponding to each subsystem contracted in some way with the $k$ upper indices corresponding to the same subsystem. These possible contractions are labeled by elements of the symmetric group $\mathfrak{S}_k$ describing permutations on $k$ elements. For example, a permutation $(1,2,3) \to (\pi(1),\pi(2),\pi(3))$ in the case of three density matrices would correspond to contracting the lower index on the first $\rho$ with the upper index on the $\pi(1)$st $\rho$, etc... . Thus, we have a basis-independent quantity for each $n$-tuple of elements of the permutation group. Specifically, we define
\be
\label{GenEnt}
S_k(\rho ; \pi_1, \cdots, \pi_n) = \rho^{a^1_1}_{a_1^{\pi_1(1)}} {}^{a^1_2}_{a_2^{\pi_2(1)}} {}^{\cdots}_{\cdots}  {}^{a^1_n}_{a_n^{\pi_n(1)}} \; \rho^{a^2_1}_{a_1^{\pi_1(2)}} {}^{a^2_2}_{a_2^{\pi_2(2)}} {}^{\cdots}_{\cdots}  {}^{a^2_n}_{a_n^{\pi_n(2)}}\cdots \rho^{a^k_1}_{a_1^{\pi_1(k)}} {}^{a^k_2}_{a_2^{\pi_2(k)}} {}^{\cdots}_{\cdots}  {}^{a^k_n}_{a_n^{\pi_n(k)}} \; .
\ee
These quantities can be represented diagrammatically via tensor networks built from the $\rho$ tensors, as in figure \ref{rho123}.

As special cases, we note that for a subsystem $A$, setting $\pi_i$ to the identity element for $i \in \bar{A}$ and $\pi_i$ to be the same cyclic permutation for each $i \in A$ gives $\tr(\rho^k)$, as shown in figure \ref{rho123}b, so the R\'enyi entropies arise as a special case. But it is not hard to check that more general quantities, such as $S_3(\rho ; (1)(23),(12)(3))$ shown in figure \ref{rho123}c, cannot generally be expressed in terms of the R\'enyi entropies and so give additional basis-independent information.\footnote{Here, the notation $(12)(3)$ indicates a permutation for which $(\pi(1), \pi(2),\pi(3)) = (2,1,3)$.} In fact, we can argue that they are the most general basis-independent quantities: each $\rho$ transforms in the adjoint representation of the unitary group realizing the change-of-basis transformations for a particular subsystem. The most general invariants built from an object in the adjoint representation (viewed as a matrix) correspond to products of traces of products of the matrix i.e. quantities with all upper indices contracted to lower indices. Thus, any invariant will involve $k$ copies of $\rho$ with all the $k$ lower indices for each subsystem contracted somehow with the $k$ upper indices for the same subsystem. Our expression (\ref{GenEnt}) represents the most general such quantity. As with the R\'enyi entropies, these quantities will not all be independent; this is clear since there can be only a finite number of independent invariant quantities when the subsystems are all finite-dimensional Hilbert spaces.

The information contained in these additional invariants which is not already present in the Renyi entropies captures the structure of multipartite entanglement in the final state. Familiar measures of  multipartite entanglement, such as the 3-tangle for three qubit systems, can be shown to be special cases of these general invariants. Like the R\'enyi entropies, the more general invariants can be related to a spectrum of eigenvalues. For the R\'enyi entropies, these are the eigenvalues of operators obtained by contracting some of the indices on a single copy of $\rho$. The more general observables can be related to eigenvalues of operators mapping subsystems to themselves obtained by contracting some of the indices starting with multiple copies of $\rho$.

\section{Calculation of average entanglement invariants}

In this section we consider the calculation of the entanglement invariants of the previous section for the final state of the black hole radiation in the general tensor network models of section 2.

We are interested in the behavior of these invariants for a generic initial black hole state. For relatively simple quantities, we expect this to agree with the average over all black hole states in the given subspace (i.e. over the microcanonical ensemble of states with a given energy). Thus, we aim to calculate the quantities (\ref{GenEnt}) averaged over the initial black hole state. We can use techniques similar to those used in the recent paper \cite{RTtensors}.

In the general expression for $S_k(\rho ; \pi_1, \cdots , \pi_n)$, the initial state of the black hole appears $k$ times as a bra and $k$ times as a ket. In calculating the average over states, we can use
\be
\int [d \Psi] \Psi_{\alpha_1} \cdots \Psi_{\alpha_k} (\Psi^\dagger)^{\beta_1}  \cdots (\Psi^\dagger)^{\beta_k} = {\Gamma(d^n) \over \Gamma(d^n+k)} \sum_{\pi \in \mathfrak{S}_k} \delta_{\alpha_1}^{\beta_{\pi(1)}} \cdots \delta_{\alpha_k}^{\beta_{\pi(k)}}
\ee
where $[d \Psi]$ is the measure on the $(2 d^{n} - 2)$-dimensional space of normalized states invariant under unitary transformations and normalized so that $\int [d \Psi] = 1$.\footnote{Equivalently, the integral over $\Psi$ can be defined as the integral over a unitary matrix $U$ with the Haar measure of an integrand obtained making the replacement $|\Psi \rangle = U |\Psi_0 \rangle$ for an arbitrary reference state $|\Psi_0 \rangle$. The result follows from standard formulae for unitary matrix integrals.} Thus, averaging over the initial state gives the normalization factor $\Gamma(d^n)/\Gamma(d^n+k) = (d^n-1)!/(d^n+k-1)!$ times the sum over ways of pairing $\Psi$s with $\Psi^\dagger$s with the replacement
\be
\Psi_\alpha (\Psi^\dagger)^\beta \to \delta_\alpha^\beta \; .
\ee
These pairings may be labeled by a permutation $\pi \in \mathfrak{S}_k$.

In the expression (\ref{GenEnt}) for a general invariant, the density matrices all take the form
\be
\label{rhoIPsi}
\rho  = {\cal I}  \Psi_{BH} \Psi_{BH}^\dagger  {\cal I}^\dagger \; .
\ee
where ${\cal I}$ is a $D^n$ times $d^n$ isometry matrix
\be
{\cal I}= ({\cal I}_1 \otimes \identity_{D^{n-1}}) ({\cal I}_2 \otimes \identity_{D^{n-2}}) \cdots ({\cal I}_{n-1} \otimes \identity_D)( {\cal I}_n)
\ee
that combines all of the isometries (\ref{defisom}) defining the model (figure \ref{fig_TN}).

\begin{figure}
\centering
\includegraphics[width=\textwidth]{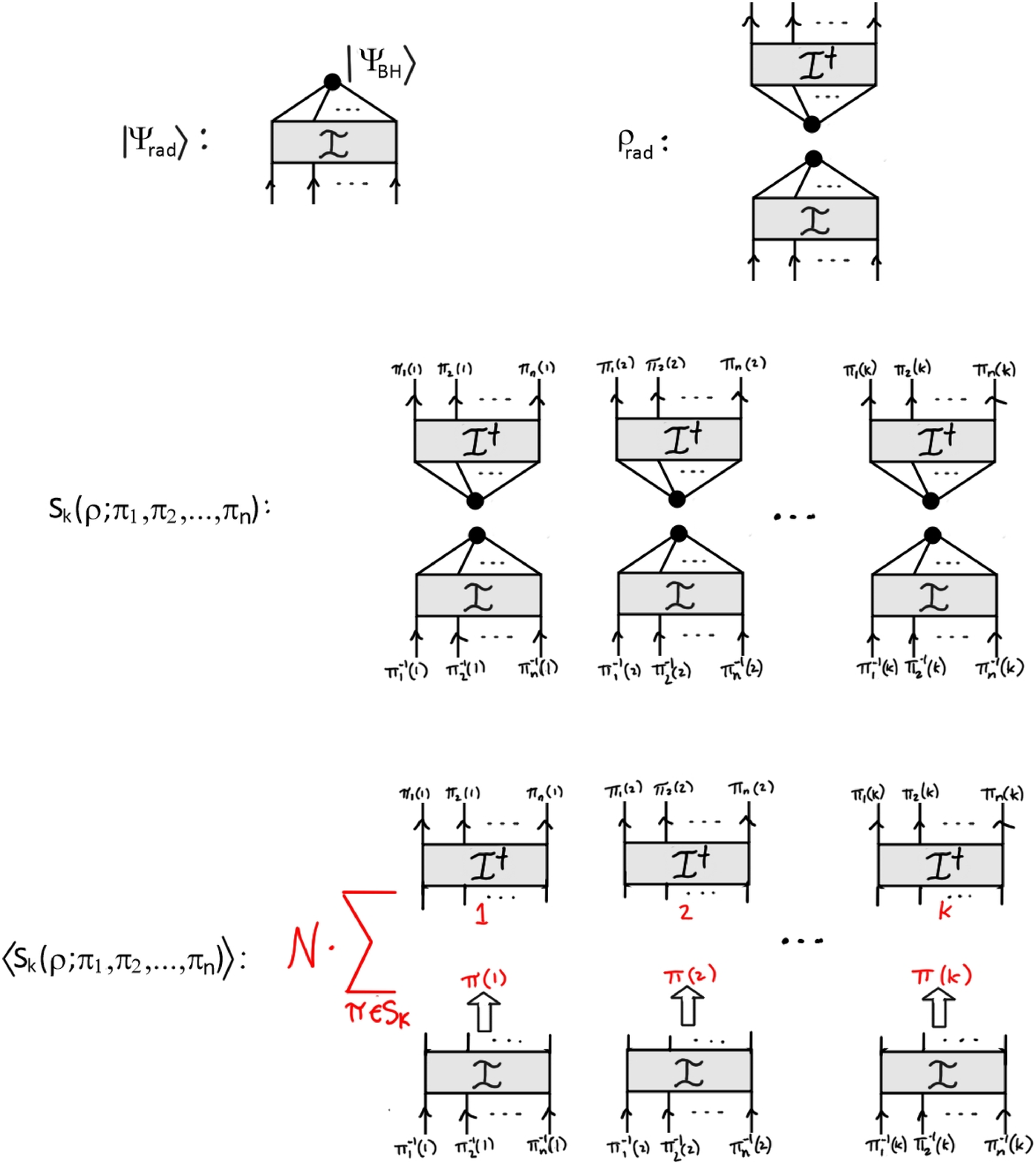}
\caption{Diagrammatic representation of $|\Psi_{rad} \rangle$ (simplified from figure \ref{fig_TN}), $\rho_{rad} = |\Psi_{rad} \rangle \langle \Psi_{rad}|$, the general invariant $S_k(\rho ; \pi_1, \cdots, \pi_n)$, and the same invariant averaged over initial black hole states. The normalization factor is ${\cal N} = \Gamma(d^n)/\Gamma(d^n + k)$.}
\label{rhocalcs}
\end{figure}

Thus, the $k$ pairings yield $k$ copies of the $D^n$ times $D^n$ matrix ${\cal I} {\cal I}^\dagger$, which is a projection matrix satisfying $({\cal I} {\cal I}^\dagger)^2 = {\cal I} {\cal I}^\dagger$. For the pairing of $\Psi$s and $\Psi^\dagger$s labeled by $\pi$, the $k$ matrices $({\cal I} {\cal I}^\dagger)_{a_1 a_2 \cdots a_n}^{b_1 b_2 \cdots b_n}$ have indices contracted according to the permutations $(\pi \cdot \pi_1, \pi \cdot \pi_2, \dots  \pi \cdot \pi_n)$, as illustrated in figure \ref{rhocalcs}.

Thus we have the general result that for the model specified by isometries ${\cal I}_n$, the averaged general invariants (\ref{GenEnt}) capturing the entanglement structure of the final radiation state are given by
\be
\label{gen_result}
\langle S_k(\rho^{rad} ; \pi_1, \cdots , \pi_n) \rangle = \prod_{m =1}^{k-1} \left(1 + {m \over d^n}\right)^{-1} \sum_{\pi \in \mathfrak{S}_k} S_k(\rho^{\cal I} ; \pi \cdot \pi_1, \cdots , \pi \cdot \pi_n)
\ee
where we have defined
\be
\rho^{\cal I} = {1 \over d^n} {\cal I} {\cal I}^\dagger \; ,
\ee
which has all the properties of a density matrix, with the additional property that $(\rho^{\cal I})^2 = {1 \over d^n} \rho^{\cal I}$. It defines a particular mixed state of the final radiation system with the property that all averaged entanglement invariants in the original model can be calculated as linear combinations of entanglement invariants for this particular state. Thus, we can think of it as a ``master state'' associated to the model.

The simplest nontrivial quantity corresponds to the case $k=2$, where we choose $\pi_i$ to be the identity for a subset $A$ of the radiation subsystems and $\pi_i$ equal to the swap permutation for the remaining subsystems, giving $\tr(\rho_A^2)$. From (\ref{gen_result}), we obtain
\bea
\label{trsquared}
\langle \tr((\rho^{rad}_A)^2) \rangle &=& S_2(\rho^{rad}_{A \bar{A}} ; id, cyc) \cr
&=& \left(1 + {1 \over d^n}\right)^{-1}  (S_2(\rho^{\cal I}_{A \bar{A}} ; id, cyc) + S_2(\rho^{\cal I}_{A \bar{A}} ; cyc, id))\cr
&=& \left(1 + {1 \over d^n}\right)^{-1} \left\{\tr((\rho^{\cal I}_A)^2) + \tr((\rho^{\cal I}_{\bar{A}})^2) \right\}
\eea
where $cyc$ is the permutation swapping the two elements. More generally, we have that
\be
\langle \tr((\rho^{rad}_A)^k) \rangle = \prod_{m =1}^{k-1} \left(1 + {m \over d^n}\right)^{-1}\sum_{\pi \in \mathfrak{S}_k} S_k(\rho^{\cal I}_{A \bar{A}} ; \pi, \pi \cdot cyc)
\ee
where in the sum, all indices corresponding to the subsystem $A$ are contracted according to $\pi$ while all indices corresponding to $\bar{A}$ are contracted according to $\pi \cdot cyc$ where $cyc$ is a cyclic permutation. Thus, in general the traces appearing in the R\'enyi entropies for subsystems of the radiation system are computed as linear combinations of the more general invariants for the auxiliary density matrix $\rho^{\cal I}$.

\section{Results for the maximally efficient models}

For the special case $D=d$ where the dimension of the radiation subsystem in which the information about the black hole initial state matches the dimension of the space of initial black hole states, the isometry ${\cal I}$ represents a unitary map, so we have that ${\cal I} {\cal I}^\dagger = \identity$ and
\be
\rho^{\cal I} = {1 \over d^n} \identity \; .
\ee
Thus, the averaged general invariants (\ref{GenEnt}) are identical in all such models, regardless of our choice of ${\cal I}_n$.

These universal results are
\beas
\langle S_k(\rho^{rad} ; \pi_1, \cdots , \pi_n) \rangle &=& \prod_{m =1}^{k-1} \left(1 + {m \over d^n}\right)^{-1} \sum_{\pi \in \mathfrak{S}_k} S_k({1 \over d^n} \identity ; \pi \cdot \pi_1, \cdots , \pi \cdot \pi_n) \cr
&=& \prod_{m =1}^{k-1} \left(1 + {m \over d^n}\right)^{-1} \sum_{\pi \in \mathfrak{S}_k} S_k(\otimes^n {1 \over d}\identity ; \pi \cdot \pi_1, \cdots , \pi \cdot \pi_n) \cr
&=& \prod_{m =1}^{k-1} \left(1 + {m \over d^n}\right)^{-1} \sum_{\pi \in \mathfrak{S}_k} S_k({1 \over d}\identity ; \pi \cdot \pi_1)  \cdots S_k({1 \over d}\identity ; \pi \cdot \pi_n) \cr
&=& \prod_{m =1}^{k-1} \left(1 + {m \over d^n}\right)^{-1} \sum_{\pi \in \mathfrak{S}_k} {d^{C(\pi \cdot \pi_1)} \over d^k}  \cdots {d^{C(\pi \cdot \pi_n)} \over d^k} \cr \; .
\eeas
Here, $C(\pi)$ is defined to be the number of cycles in the permutation $\pi$, and to obtain the third line, we have used that
\be
S_k(\rho_1 \otimes \cdots \otimes \rho_n, ; \pi_1, \cdots , \pi_n) = S_k(\rho_1 ; \pi_1) \cdots S_k(\rho_n ; \pi_n)
\ee
which follows from the definition. The last line follows since $S_k(\identity ; \pi)$ corresponds to taking $k$ copies of the identity matrix and contracting according to $\pi$ which gives a product of traces of the $d \times d$ identity matrix, one for each cycle in $\pi$. Rearranging the last line above, we obtain the final result
\be
\label{gen_result_U}
\langle S_k(\rho^{rad} ; \pi_1, \cdots , \pi_n) \rangle = {\Gamma(d^n) \over \Gamma(d^n + k)} \sum_{\pi \in \mathfrak{S}_k} d^{C(\pi \cdot \pi_1)+ \dots + C(\pi \cdot \pi_n)} \; ,
\ee
or equivalently
\be
\label{gen_result_U1}
\langle S_k(\rho^{rad} ; \pi_1, \cdots , \pi_n) \rangle = {\Gamma(d^n) \over \Gamma(d^n + k)} \sum_{\pi \in \mathfrak{S}_k}  \prod_{i} D_i^{C(\pi \cdot \pi_i)} \; ,
\ee
where $D_i$ represents the total dimension of the subsystem whose indices are contracted according to the permutation $\pi_i$.

Diagrammatically, this result is obtained from the last image in figure \ref{rhocalcs} by removing all the ${\cal I}$s and ${\cal I}^\dagger$s, since in each case, these contract with each other to give an identity matrix. We then connect up all the lines according to the permutations, and assign a factor of $d$ to each loop, since the loops each correspond to a trace of the $d \times d$ identity matrix. The calculation in figure \ref{rhocalcs} with the isometries amounts to calculating the entanglement invariants for a general state of the radiation system and then averaging over all possible states $|\Psi \rangle = U |\Psi_0 \rangle$ via the Haar measure on unitary matrices $U$. Thus, the universal results that we obtain in this case are exactly the entanglement invariants in Haar-random multipartite systems. Some of these were previously known and can be used to provide a check of our result; in other cases, our formulae give new results for general entanglement invariants of Haar-random systems.

As examples, we can calculate the average for the quantities shown in figure \ref{rho123} where we take the subsystem $B$ to represent the first $j$ radiation subsystems (i.e. the systems storing the first $j$ quanta emitted). From the formula (\ref{gen_result_U}), we have for the quantity in figure \ref{rho123}b,
\beas
\langle \tr((\rho^{rad}_j)^2) \rangle &=& \frac{1}{d^n(d^n+1)} \left( d^{(n-j)C(id \cdot id) + j C(id \cdot cyc)} + d^{(n-j)C(cyc \cdot id) + j C(cyc \cdot cyc)}  \right) \cr
&=& \frac{1}{d^n(d^n+1)} (d^{2n-j} + d^{n + j}).
\label{eq:avgTrRhoSq}
\eeas
Using this result, we can obtain a measure of how close the density matrix describing the first $j$ radiated quanta is to the maximally mixed density matrix for that system. Making use of the dstance measure ${\rm Dist}(\rho,\sigma) = \sqrt{\tr((\rho-\sigma)^2)}$ we obtain from the previous result that the averaged distance to the maximally mixed state is
\be
\langle {\rm Dist}^2 \rangle = \langle \tr\left( \left(\rho_j - {1 \over d^j} \identity \right)^2 \right) \rangle = \langle \tr((\rho^{rad}_j)^2) \rangle - {1 \over d^j} = {1 \over d^n + 1} \left[ d^j - {1 \over d^j} \right] \; .
\ee
Recalling that $n$ represents the entropy of the black hole, we see that for $j \ll n$, the density matrix for the radiation subsystem is extremely close to maximally mixed, of order $e^{- c S_{BH}}$ for some order one number $c$. Only when $j$ is nearly equal to $n$ (i.e. almost all quanta emitted) is the radiation density matrix significantly different from maximally mixed; this agrees with previous expectations \cite{Page, HayPres, Hay2, Zyc1}.

As another simple example, we can compute the average for the more general quantity shown in figure \ref{rho123}c. Denoting the elements of $\mathfrak{S}_3$ by
\be
\{id = (1)(2)(3), \sigma_{12} = (12)(3), \sigma_{23} = (1)(23), \sigma_{13} = (2)(13), cyc = (123),cyc^{-1} = (132)\} \; .
\ee
we have
\be
C(id) = 3 \qquad C(\sigma_{12}) = C(\sigma_{23}) = C(\sigma_{13}) = 2 \qquad C(cyc) = C(cyc^{-1}) = 1 \; .
\ee
Thus, we obtain from (\ref{gen_result_U}), using the multiplication rules for the permutation group,
\be
S_3(\rho^{rad}_{A \bar{A}} ; \sigma_{23}, \sigma_{12}) =  \frac{d^{3n - 2j} + 3 d^{2n} + d^{n + 2j} + d^n}{d^n(d^n+1)(d^n + 2)} \; .
\ee
Here again, we take the subsystem $A$ to be the first $j$ radiation subsystems.

\subsubsection*{Result for general invariants in a bipartite system}

Calculating these invariants directly as in our examples becomes cumbersome as $k$ becomes larger. Fortunately, we can come up with a somewhat more explicit formula in the case where only two different permutations matrices appear; this includes as a special case the R\'enyi entopies for arbitrary subsystems.

Starting from the general expression \ref{gen_result_U1}, we have
\beas
\langle S_3(\rho^{rad}_{A \bar{A}} ; \pi_1, \pi_2) \rangle &=& {\Gamma(d^n) \over \Gamma(d^n + k)} \sum_{\pi \in \mathfrak{S}_k}  \prod_{i} d^{j \; C(\pi \cdot \pi_1) + (n-j) \; C(\pi \cdot \pi_2)} \cr
&=& {\Gamma(d^n) \over \Gamma(d^n + k)} \sum_{\sigma \in \mathfrak{S}_k}  \prod_{i} d^{j \; C(\sigma) + (n-j) \; C(\sigma \cdot \pi_1^{-1} \pi_2)}
\eeas
where we have changed the summation variable to $\sigma = \pi \pi_1$. Thus, the result depends only on the combination $\pi_1^{-1} \pi_2$.

Using the methods of enumerative combinatorics, we can express this directly in terms of the irreducible characters $\chi^{\lambda} : \mathfrak{S}_k \rightarrow \mathbb{C}$ with $\lambda$ a partition of $n$ giving the cycle structure of the permutation. The result, derived in detail in the appendix, is:
\begin{equation}
	\langle S_3(\rho^{rad}_{A \bar{A}} ; \pi_1, \pi_2) \rangle = \frac{\Gamma(d^n)}{k!\Gamma(d^n + k)} \sum_{\lambda \vdash k}  f^{\lambda} \prod_{u \in \lambda} (d^n+c(u))(d^{n-j}+c(u)) \chi^{\lambda}(\pi_1^{-1} \pi_2),
\label{eq:combQ}
\end{equation}
where the sum is over all partitions $\lambda$ of $k$ and $c : Par(k) \rightarrow \mathbb{Z}$ gives the content of the $u^{th}$ cell in the Young diagram of $\lambda$.

We can apply this to the case where $\pi_1 = id$ and $\pi_2 = (1~2~\cdots~n)$ to obtain an expression for the trace of the $k$th power of the reduced density matrix for the subsystem with $j$ parts. As shown in the appendix, we obtain:
\begin{equation}
  \begin{gathered}
	\langle Tr((\rho_A)^k) \rangle = \frac{(d^n-1)!}{k (d^n+k-1)!} \sum_{l=0}^{k-1} \frac{(-1)^l}{l! (k-l-1)!} \frac{(d^{n-j}+k-l-1)! (d^j+k-l-1)!}{(d^{n-j}-l-1)! (d^j-l-1)!} \\
	= \frac{\Gamma(d^n)}{k \Gamma(d^n+k)} \sum_{l=0}^{k-1} \frac{(-1)^l}{l! \Gamma(k-l)} \frac{\Gamma(d^{n-j}+k-l) \Gamma(d^j+k-l)}{\Gamma(d^{n-j}-l) \Gamma(d^j-l)} \\
	= \frac{\Gamma(k+d^{n-j})\Gamma(k+d^j)\Gamma(d^n)}{k \Gamma(d^n+k)\Gamma(k)\Gamma(d^{n-j})\Gamma(d^j)}~ {}_3F_2 \bigg(
	\begin{matrix}
	1-k, 1-d^{n-j}, 1-d^j;\\
	1-k-d^{n-j}, 1-k-d^j;
	\end{matrix}
	1 \bigg).
  \end{gathered}
\label{eq:trRhon}
\end{equation}
where the last line is a generalized hypergeometric function. Via (\ref{renyi}) this gives a general result for the Renyi entropies of arbitrary subsystems of the final radiation system. This result has been found to agree with the formula for Renyi entropies of Haar-random states previously obtained in \cite{Malacarne} (for specific powers up to 10) though our result gives a more explicit closed form.

The derivation of equation \ref{eq:trRhon} relied only on the factorization of the Hilbert space into a bipartite system, thus, this formula holds for the reduced density matrix of an $m$ dimensional subsystem A in any bipartite Hilbert space of dimension $mp$ simply by replacing $d^j,~d^{n-j}$, and $d^n$ by $m,~p$, and $mp$ respectively. Finally, using the formula
\begin{equation}
	S = -Tr(\rho log(\rho)) = -\frac{d}{dn}\langle Tr(\rho^n) \rangle |_{n=1} = \lim_{n \to 1} S_n,
\label{eq:StrRhon}
\end{equation}
for entanglement entropy in terms of Renyi entropies, we obtain (taking $j \leq n-j$)
\begin{equation}
	S = \sum_{q=d^{n-j}+1}^{d^n} \frac{1}{q} - \frac{d^{2j}-d^j}{2d^n},
\label{eq:PageEntropy}
\end{equation}
which is precisely Page's formula for the average entanglement entropy of a subsystem in a bipartite Hilbert space \cite{Page}. This provides a check of our results.

\section{Constraints from an ``information-free'' horizon}

For the case $D = d$, we have seen that all entanglement invariants are independent of the specific model once averaged over all possible initial black hole states. This will not be the case in the more general situation where $D > d$. In this case, we can try to come up with constraints on the isometries ${\cal I}$ based on information-theoretic quantities in order to incorporate additional physics expected in black hole systems.

A characteristic feature of black hole systems is that the horizon is ``information-free'' within the semiclassical approximation. In Hawking's calculation, the black hole radiation is completely independent of the microstate of the black hole apart from its overall conserved charges. In a unitary model, this cannot be precisely true, but we can try to demand that the outgoing radiation carries as little information as possible, on average, about the black hole state.

As an example of a specific criterion, we can try to choose isometries ${\cal I}_n$ so that the state of the radiation subsystem after an emission step (\ref{defisom}) is as close as possible to being maximally mixed. Specifically, we will try to minimize
\be
\label{dist1}
\langle {\rm Dist}^2(\rho_{rad}, \identity) \rangle = \langle \tr((\rho_{rad} - {1 \over D} \identity)^2) \rangle = \langle \tr(\rho_{rad}^2) \rangle - {1 \over D}
\ee
where the average is over all initial states. Here, the particular choice of distance measure is for technical convenience.

For each $n$, the isometry is a map from a $d^n$ dimensional Hilbert space to a tensor product Hilbert space with factors of dimension $d^{n-1}$ and $D$. This is described explicitly by a tensor ${\cal I}^a_{b c}$ where we are suppressing the label $n$. It will be convenient to associate this to a pure state
\be
|\hat{\Psi} \rangle = \sum_{a,b,c} {\cal I}^a_{b c} |a \rangle \otimes |b \rangle \otimes |c \rangle \; .
\ee
of a tripartite system, whose parts we will label as $A,B,C$.

Using the result (\ref{trsquared}) we find that in terms of the state $|\Psi_{\cal I} \rangle$, the averaged distance in (\ref{dist1}) is given by
\be
\label{minrho}
\langle {\rm Dist}^2(\rho_{rad}, \identity) \rangle \propto \tr(\hat{\rho}_B^2) + \tr(\hat{\rho}_C^2) \; .
\ee

As $\hat{\rho}_B$ and $\hat{\rho}_C$ are reduced density matrices, they are hermitian and have trace one. As a simple application of the Cauchy-Schwarz inequality, one can show that a $N \times N$ hermitian matrix $M$ with $tr(M)=Q$ must satisfy $tr(M^2) \geq Q^2/N$, with equality if and only if $M \propto \identity_N$. Hence, one can minimize equation \ref{minrho} by choosing ${\cal I}$ so that $\hat{\rho}_B$ and $\hat{\rho}_C$ are both proportional to identity. Further, the condition that ${\cal I}$ is an isometry is precisely that $\hat{\rho}_A \propto \identity_{d^n}$. Thus, the radiation will be as close as possible to maximally mixed if the pure state associated with each isometry ${\cal I}_n$ has each subsystem maximally mixed.

Isometries achieving this condition of corresponding to states with maximally mixed subsystems can exist only for certain dimensions $D$ which are not too large relative to $d$. By the Schmidt decomposition, each subsystem's reduced density matrix must have the same non-zero eigenvalues as the density matrix describing the other two subsystems together. Since we require these matrices to be full rank, the dimension of each subsystem must be less than or equal to the product of the other two subsystems' dimensions. So we have that $d \le D \le d^{2n - 1}$ in order to achieve the minimum suggested by the Cauchy-Schwarz inequality.\footnote{For larger $D$, the minimum is that for $D = d^{2n - 1}$ and can be achieved by taking the isometry to map into a dimension $d^{2n - 1}$ subsystem of the radiation system}.

Even for $d \le D \le d^{2n - 1}$ it is not always possible to find pure states with maximally mixed subsystems. It turns out that the problem of constructing and classifying such states for general dimensions $(A,B,C)$ is a rich mathematical question with connections to symplectic geometry, geometric invariant theory, and representation theory of finite and compact groups. On the quantum information theory side, it is related to the problem of classifying states under LOCC (local operations and classical communication) transformations. A more complete discussion of this interesting problem will be the subject of an upcoming paper \cite{toappear}.

\section{Discussion}

In this paper, we have introduced technical tools based on tensor networks for studying a broad class of toy models for unitary black hole evaporation. Given a set of isometries specifying any such model, averages for general bipartite/multipartite entanglement measures may be computed directly from the general result (\ref{gen_result}). Applying this in the case where the dimension of the final radiation system is the same as the initial black hole subsystem, we have obtained explicit results that are model-independent. We have argued that these must match the results for multipartite entanglement measures in Haar random systems, and show that our expressions agree with and generalize known formulae for this case.

Using our technology, it would be interesting to investigate specific models incorporating additional features of black hole physics. We have explored one such direction, finding conditions on the isometries which guarantee that the radiation state after an evolution step is as close to maximally mixed as possible. Interestingly, the required isometry tensors have the property that the associated quantum states have each subsystem maximally mixed. This same property appears for perfect tensors, which played a key role in another recent toy model of a gravitational system, the HaPPY model of holographic states \cite{Happy}. It would be interesting to further explore our black hole models built from tensors with these properties and to consider other possible conditions designed to capture in more detail the physics of black holes.

\section*{Acknowledgements}

We would like to thank Alex May for very useful discussions in the early stages of this work and Richard Stanley (via Math Overflow) for pointing out the usefulness of reference \cite{Stanley}. This research is supported in part by the Natural Sciences and Engineering Research Council of Canada, and by grant 376206 from the Simons Foundation.

\appendix

\section{Combinatoric Details of the Derivation}

In this appendix, we provide the derivations for (\ref{eq:trRhon}) and (\ref{eq:combQ}) in section 5. This derivation entirely relies on results from ``Enumerative Combinatorics" (volume 2) by Richard Stanley and thus, specific results from this reference are stated without proof (although all proofs can be found in \cite{Stanley}).

First, we note the following result relating the product of three Schur functions (denoted by $s_{\lambda}$) to the product of three poly-sum symmetric functions (denoted by $p_{\rho}$):
\begin{equation}
	\sum_{\lambda \vdash k} \bigg(\prod_{u \in \lambda} h(u) \bigg) s_{\lambda} s_{\lambda} s_{\lambda} = \frac{1}{k!} \sum_{\omega_1 \omega_2 \omega_3 = id} p_{\rho(\omega_1)}p_{\rho(\omega_2)}p_{\rho(\omega_3)},
\label{eq:6} 
\end{equation}
where $h : Par(k) \rightarrow \mathbb{Z}$ gives the hook length of the $u^{th}$ cell in the Young diagram of $\lambda$, $\omega_1, \omega_2, \omega_3 \in \mathfrak{S}_k$ and $\rho : \mathfrak{S}_k \rightarrow Par(k)$ gives the cycle structure of the permutation as a partition of k. Furthermore, each symmetric polynomial on either side of equation \ref{eq:7} is evaluated at a distinct set of indeterminates, with the corresponding symmetric polynomial on the other side evaluated at the same set of indeterminates. Setting the first set of indeterminates to consist of $x$ ones with the rest zero and the second set to consist of $y$ ones with the rest zero (the third set is left unspecified), we obtain, using corollary 7.21.4 of \cite{Stanley} and the definition of power-sum symmetric polynomials:
\begin{equation}
	\sum_{\lambda \vdash k} \bigg(\prod_{u \in \lambda} h(u) \bigg) \prod_{u \in \lambda} \frac{x+c(u)}{h(u)} \prod_{u \in \lambda} \frac{y+c(u)}{h(u)} s_{\lambda} = \frac{1}{k!} \sum_{\omega_1 \omega_2 \omega_3 = id} x^{d(\omega_1)}y^{d(\omega_2)} p_{\rho(\omega_3)},
\label{eq:7}  
\end{equation}
where $d : \mathfrak{S}_k \rightarrow \mathbb{N}$ gives the number of cycles in the decomposition of the permutation and $c : Par(k) \rightarrow \mathbb{Z}$ gives the content of the $u^{th}$ cell in the Young diagram of $\lambda$. Cancelling the hook length products on the left-hand side (LHS) of equation \ref{eq:7} and multiplying by $k!$ gives, by corollary 7.21.6 of \cite{Stanley}:
\begin{equation}
	\sum_{\lambda \vdash k}  f^{\lambda} \prod_{u \in \lambda} (x+c(u))(y+c(u)) s_{\lambda} = \sum_{\omega_1 \omega_2 \omega_3 = id} x^{d(\omega_1)}y^{d(\omega_2)} p_{\rho(\omega_3)},
\label{eq:8} 
\end{equation}
where
\begin{equation}
f^{\lambda} = \frac{n!}{\prod_{u \in \lambda} h(u)}.
\label{eq:flambda}
\end{equation}
We define the polynomial $F_{\lambda} (x,y)$:
\begin{equation}
	F_{\lambda} (x, y) = \sum_{\pi \in \mathfrak{S}_k} x^{d(\pi)} y^{d(\pi \pi_1)},
\label{eq:9} 
\end{equation}
where $\rho(\pi_1) = \lambda$, as this exact quantity (for $\pi_1 = (1~2 \cdots k)$) appears in the expression for $\langle Tr(\rho^k) \rangle$. We also define the following symmetric polynomial:
\begin{equation}
	G_k = \sum_{\lambda \vdash k} \frac{k!}{z_{\lambda}} F_{\lambda} (x,y) p_{\lambda}
	        = \sum_{\sigma \in \mathfrak{S}_k} \sum_{\pi \in \mathfrak{S}_k} x^{d(\pi)} y^{d(\pi \sigma)} p_{\rho(\sigma)},
\label{eq:10} 
\end{equation}
where
\begin{equation}
z_{\lambda} = \prod_{j=1}^k (j)^{a_j} (a_j)!,
\label{zlambda}
\end{equation}
with $a_j$ giving the number of times $j$ appears in $\lambda$. The second equality of (\ref{eq:10}) follows as $\frac{k!}{z_{\lambda}}$ is the number of elements in the conjugacy class given by $\lambda$ and $F_{\lambda} (x,y)$ is invariant under conjugation of $\pi_1$. We now show that $G_k$ is equal to the right-hand side (RHS) of equation \ref{eq:8}. Multiplying by $\omega_3^{-1}$ on right of the summation index in equation \ref{eq:8} we obtain $\omega_1 \omega_2 = \omega^{-1}_3$ as an equivalent representation of the summation index. For any given $\omega_1$ and $\omega_3$, there is a unique choice of $\omega_2$ such that the summation index equation holds. Thus, for each choice of $\omega_3$, there are exactly $k! = |\mathfrak{S}_k|$ pairs $(\omega_1, \omega_2) \in (\mathfrak{S}_k)^2$ (exactly one $\omega_2$ for each choice of $\omega_1$) that contribute terms to the sum involving that particular $\omega_3$. Furthermore, the sum runs over all $\omega_3 \in \mathfrak{S}_k$, giving a total of $\frac{k!}{z_{\lambda}} k!$ terms in the sum involving $p_{\lambda}$. Looking at equation \ref{eq:10}, these two expressions clearly involve the same of number of terms with $p_{\lambda}$ for any given cycle type $\lambda$, thus, it remains to show that, in these sums, the powers to which x and y are raised are the same.

First, let $\sim$ be the conjugacy equivalence relation in $\mathfrak{S}_k$ (i.e. for $\pi, \sigma \in \mathfrak{S}_k$: $\pi \sim \sigma \Leftrightarrow \exists \tau \in \mathfrak{S}_k$ s.t. $\tau \sigma \tau^{-1} = \pi \Leftrightarrow \pi$ and $\sigma$ have the same cycle structure). Obviously if $\pi \sim \sigma$, then $\rho(\pi) = \rho(\sigma)$ and $d(\pi) = d(\sigma)$. Now suppose we fix $\sigma$ of equation \ref{eq:10} and $\omega_3 \sim \omega_3^{-1}$ of equation \ref{eq:8} so that they are conjugate in $\mathfrak{S}_k$. Then we have $\tau \in \mathfrak{S}_k$ such that $\tau \omega_3^{-1} \tau^{-1} = \sigma$. Using the summation index expression $\omega_2 = \omega_1^{-1} \omega_3^{-1}$, and conjugating by $\tau$, we obtain $\tau \omega_2 \tau^{-1} = \tau \omega_1^{-1} \tau^{-1} \tau \omega_3^{-1} \tau^{-1} = \pi' \sigma$, where $\pi' \sim \omega_1^{-1} \sim \omega_1$. Thus for some $\pi' \in \mathfrak{S}_k$ conjugate to $\omega_1$, $\omega_2$ is conjugate to $\pi' \sigma$, when $\omega_3 \sim \sigma$. So $x^{d(\pi')} y^{d(\pi' \sigma)} = x^{d(\omega_1)} y^{d(\omega_2)}$, and since we are summing over all of $\mathfrak{S}_k$ in both equations \ref{eq:8} and \ref{eq:10}, every $(\omega_1, \omega_2)$ in equation \ref{eq:8} is matched by a pair $(\pi', \pi' \sigma)$ in equation \ref{eq:10} with $\rho(\omega_3) = \rho(\sigma)$. Thus equation \ref{eq:8} gives another formula for $G_k$.

Finally, applying theorem 7.18.5 of \cite{Stanley} to equation \ref{eq:8}, we obtain:
\begin{equation}
	G_k = \sum_{\lambda \vdash k}  f^{\lambda} \prod_{u \in \lambda} (x+c(u))(y+c(u)) \sum_{\mu \vdash \lambda} z_{\mu}^{-1} \chi^{\lambda}(\mu)p_{\mu},
\label{eq:11} 
\end{equation}
where $\chi$ is an irreducible character of $\mathfrak{S}_k$. Proposition 7.9.3 of \cite{Stanley} gives the scalar product of power-sum symmetric polynomials, thus, taking the scalar product of $G_k$ with $p_{\nu}$, we obtain:
\begin{equation}
  \begin{gathered}
	\langle p_{\nu}, G_k \rangle = \sum_{\lambda \vdash k}  f^{\lambda} \prod_{u \in \lambda} (x+c(u))(y+c(u)) \sum_{\mu \vdash \lambda} z_{\mu}^{-1} \chi^{\lambda}(\mu) z_{\nu} \delta_{\nu \mu}\\
	= \sum_{\lambda \vdash k}  f^{\lambda} \prod_{u \in \lambda} (x+c(u))(y+c(u)) \chi^{\lambda}(\nu)\\
	= \sum_{\lambda \vdash k} \frac{k!}{z_{\lambda}} F_{\lambda} (x,y) z_{\nu} \delta_{\nu \lambda} = k! F_{\nu} (x,y).
  \end{gathered}\label{eq:12}  
\end{equation}
We have therefore extracted the following final result for the polynomial $F_{\lambda} (x,y)$:
\begin{equation}
	F_{\nu} (x,y) = \frac{1}{k!} \sum_{\lambda \vdash k}  f^{\lambda} \prod_{u \in \lambda} (x+c(u))(y+c(u)) \chi^{\lambda}(\nu).
\label{eq:13} 
\end{equation}
In the case of a cyclic permutation as $\pi_1$ in the definition of $F_{\lambda} (x,y)$ (equation \ref{eq:9}), we have $\rho(\pi_1) = \langle k \rangle = \nu$. In this case, the Murnaghan-Nakayama rule (Theorem 7.17.3 of \cite{Stanley}), gives the irreducible character $\chi^{\lambda}(\langle k \rangle)$ as non-zero only in the case that $\lambda$ is a partition of $k$ with at most one part greater than 1, in which case the irreducible character is $(-1)^{l(\lambda) - 1}$, where $l : Par(k) \rightarrow \mathbb{N}$ gives the number of parts of the partition. Suppose we have $\lambda = \langle 1^m k-m \rangle$, then $l(\lambda) = m + 1$, so $\chi^{\lambda}(\langle k \rangle) = (-1)^m$. From the Young diagram of $\lambda$ it can be shown that $c(u) \in \{-m, -m+1, ..., -1, 0, 1, ..., k-m-1\} = W$, with $c(u)$ taking each value in $W$ exactly once when going over all $u \in \lambda$. Also from the Young diagram, along the column $h(u) \in \{1, 2, ..., m\}$, at the corner $h(u) = k$, and along the row $h(u) \in \{1, 2, ..., k-m-1\}$, with each value taken on exactly once when going over all $u \in \lambda$. Thus, for this $\lambda \vdash k$, the term in the sum of equation \ref{eq:13} is $(-1)^m \frac{k!}{k (k-m-1)! m!} \prod_{j=-m}^{k-m-1} (x+j)(y+j)$. Now there is one partition of $k$ of this type for each $m \in \{0, 1, ..., k-1\}$, so collecting terms into factorials we have:
\begin{equation}
  \begin{gathered}
	F_{\langle k \rangle} (x,y) = \frac{1}{k!} \sum_{m=0}^{k-1} (-1)^m {k-1 \choose m, k-m-1} xy \prod_{j=1}^m (x-j)(y-j)\prod_{m=1}^{k-m-1} (x+j)(y+j)\\
	= \frac{1}{k} \sum_{m=0}^{k-1} \frac{(-1)^m}{m! (k-m-1)!} \frac{(x+k-m-1)! (y+k-m-1)!}{(x-m-1)! (y-m-1)!}.
  \end{gathered}\label{eq:14} 
\end{equation}
Equation \ref{eq:14} gives the polynomial appearing the expression for $\langle Tr(\rho^k) \rangle$ (equation \ref{eq:trRhon}), since the matrix multiplication in $\rho^k$ corresponds to index contractions represented by a cyclic permutation.


\begin{thebibliography}{99}
\bibitem{creation}
  S.~W.~Hawking,
  ``Particle creation by black holes,'' Communications in Mathematical Physics, {\bf 46}, 2 (1975)
  doi:10.1007/BF01608497
\bibitem{breakdown}
  S.~W.~Hawking,
  ``Breakdown of Predictability in Gravitational Collapse,'' Phys.\ Rev.\ D {\bf 14}, 2460 (1976).
  doi:10.1103/PhysRevD.14.2460
\bibitem{Gidd2013}
 S.~B.~Giddings,
  ``Black holes, quantum information, and unitary evolution,''
  [arXiv:1201.1037 [hep-th]]
\bibitem{MathurPed}
  S.~D.~Mathur,
  ``The Information paradox: A Pedagogical introduction,'' Class.\ Quant.\ Grav.\  {\bf 26}, 224001 (2009)
  doi:10.1088/0264-9381/26/22/224001
  [arXiv:0909.1038 [hep-th]].
\bibitem{Gidd1992}
  S.~B.~Giddings,
  ``Toy models for black hole evaporation,''
  [arXiv:hep-th/9209113]
\bibitem{Gidd2012}
  S.~B.~Giddings,
  ``Models for unitary black hole disintegration,''
  Phys.\ Rev.\ D {\bf 85}, 044038 (2012)
  doi:10.1103/PhysRevD.85.044038
  [arXiv:1108.2015 [hep-th]].\bibitem{GiddShi}
  S.~B.~Giddings and Y.~Shi,
  ``Quantum information transfer and models for black hole mechanics,'' [arXiv:1205.4732 [hep-th]]
\bibitem{MathurInfall}
  S.~D.~Mathur,
  ``The Information paradox and the infall problem,''
  Class.\ Quant.\ Grav.\  {\bf 28}, 125010 (2011)
  doi:10.1088/0264-9381/28/12/125010
  [arXiv:1012.2101 [hep-th]].
\bibitem{Avery}
  S.~G.~Avery,
  ``Qubit Models of Black Hole Evaporation,''
  JHEP {\bf 1301}, 176 (2013)
  doi:10.1007/JHEP01(2013)176
  [arXiv:1109.2911 [hep-th]].
\bibitem{Page2016}
  K.~Osuga and D.~N.~Page,
  ``Qubit Transport Model for Unitary Black Hole Evaporation without Firewalls,''
  arXiv:1607.04642 [hep-th].
\bibitem{Rozali}
  B.~Czech, K.~Larjo and M.~Rozali,
  ``Black Holes as Rubik's Cubes,'' JHEP {\bf 1108}, 143 (2011)
  doi:10.1007/JHEP08(2011)143
  [arXiv:1106.5229 [hep-th]].
\bibitem{Bradler} 
  K.~Bradler and C.~Adami,
  ``One-shot decoupling and Page curves from a dynamical model for black hole evaporation,''
  Phys.\ Rev.\ Lett.\  {\bf 116}, no. 10, 101301 (2016)
  doi:10.1103/PhysRevLett.116.101301
  [arXiv:1505.02840 [quant-ph]].  
\bibitem{Richter} 
  G.~Dvali, C.~Gomez, D.~Lust, Y.~Omar and B.~Richter,
  ``Universality of Black Hole Quantum Computing,''
  Fortsch.\ Phys.\  {\bf 65}, 46 (2017)
  doi:10.1002/prop.201600111
  [arXiv:1605.01407 [hep-th]].
  
\bibitem{Harlow}
  D.~Harlow,
  ``Jerusalem Lectures on Black Holes and Quantum Information,''
  Rev.\ Mod.\ Phys.\  {\bf 88}, 15002 (2016)
  [Rev.\ Mod.\ Phys.\  {\bf 88}, 15002 (2016)]
  doi:10.1103/RevModPhys.88.015002
  [arXiv:1409.1231 [hep-th]].
\bibitem{Page}
  D.~N.~Page,
  ``Average entropy of a subsystem,'' Phys.\ Rev.\ Lett.\  {\bf 71}, 1291 (1993)
  doi:10.1103/PhysRevLett.71.1291
  [gr-qc/9305007].
\bibitem{Malacarne}
  L.~C.~Malacarne, R.~S.~Mendes and E.~K.~Lenzi,
  ``Average entropy of a subsystem from its average Tsallis entropy,''
  Phys.\ Rev.\ E {\bf 65}, 046131 (2002).
  doi:10.1103/PhysRevE.65.046131
\bibitem{Zyc2}
  H.-J.~Sommers and K.~\.Zyczkowski,
  ``Statistical properties of random density matrices,''
  J. Phys. A: Math. Gen. {\bf 37}, 8457 (2004)
\bibitem{SLthesis}
  S.~Leutheusser,
  ``Development of a toy model for black hole evaporation,''
  University of British Columbia Undergraduate Thesis, (2016)
\bibitem{RTtensors}
  P.~Hayden, S.~Nezami, X.~L.~Qi, N.~Thomas, M.~Walter and Z.~Yang,
  ``Holographic duality from random tensor networks,''
  JHEP {\bf 1611}, 009 (2016)
  doi:10.1007/JHEP11(2016)009
  [arXiv:1601.01694 [hep-th]].
\bibitem{HayPres}
  P.~Hayden and J.~Preskill,
  ``Black holes as mirrors: Quantum information in random subsystems,''
  JHEP {\bf 0709}, 120 (2007)
  doi:10.1088/1126-6708/2007/09/120
  [arXiv:0708.4025 [hep-th]].


\bibitem{Hay2}
  P.~Hayden, D.~W.~Leung, and A.~Winter,
  ``Aspects of generic entanglement,''
  Communications in Mathematical Physics {\bf 265}, 1 (2006)
  [arXiv:quant-ph/0407049]
\bibitem{Zyc1}
  H.-J.~Sommers and K.~\.Zyczkowski,
  ``Induced measures on the space of mixed quantum states,''
  J. Phys. A: Math. Gen. {\bf 34}, 7111 (2001)
\bibitem{toappear}
 S. ~Leutheusser, A. ~May, M.~ Van Raamsdonk, ``On locally maximally mixed states of multipart quantum systems.''
\bibitem{Happy}
  F.~Pastawski, B.~Yoshida, D.~Harlow and J.~Preskill,
  ``Holographic quantum error-correcting codes: Toy models for the bulk/boundary correspondence,''
  JHEP {\bf 1506}, 149 (2015)
  doi:10.1007/JHEP06(2015)149
  [arXiv:1503.06237 [hep-th]].
\bibitem{Stanley}
  R.~P.~Stanley,
  ``Enumerative Combinatorics Volume 2''
  Cambridge, UK: Cambridge University Press (2012) 585 p
\end{thebibliography}
\end{document}